\documentstyle[aps,graphics]{revtex}

\begin{document}

\author{Guillermo Abramson \footnote{Present address: 
Max-Planck-Institut f\"ur Physik Komplexer Systeme, N\"othnitzer 
Str. 38, 01187 Dresden, Germany.} }
\address{International Centre for 
Theoretical Physics \\ \it Strada Costiera 11, P.O. Box 586, 34100 
Trieste, Italy }
\author{Dami\'{a}n H. Zanette \footnote{Permanent address: Consejo
Nacional de Investigaciones Cient\'{\i}ficas y T\'ecnicas, Centro
At\'omico Bariloche, 8400 Bariloche, Argentina.} }
\address{Fritz-Haber-Institut der Max-Planck-Gesellschaft \\ \it Faradayweg
4-6, 14195 Berlin, Germany}
\title{Statistics of extinction and survival in Lotka-Volterra systems}

\maketitle

\begin{abstract}

We analyze purely competitive many-species Lotka-Volterra systems with
random  interaction  matrices,  focusing  the attention on statistical
properties  of  their  asymptotic  states.  Generic  features  of  the
evolution  are  outlined  from  a  semiquantitative  analysis  of  the
phase-space  structure,  and   extensive  numerical  simulations   are
performed to study the statistics of the extinctions. We find that the
number  of  surviving  species  depends  strongly  on  the statistical
properties  of  the  interaction  matrix,  and that the probability of
survival is weakly correlated to specific initial conditions.

\end{abstract}
\draft


\section{Introduction}

Systems of  interacting biological  species evolve  through the  long,
slow and intricate process of natural selection \cite{evol}.  Usually,
the result of  this process is  so complex that  the dynamics of  such
webs of coevolving  species  can  be successfully represented,  within
relatively short  time scales,  by means  of a  dynamical system  with
stochastic elements \cite{may}.  A standard mathematical model for the
joint evolution of $M$  biological species with spatially  homogeneous
densities $n_i(t)$ ($i=1,2,...,M$)  is the generalized  Lotka-Volterra
system \cite{rusos},
\begin{equation}
\dot{n}_i(t)=n_i(t)\left[r_i-\sum_{j=1}^M\kappa _{ij}n_j(t)\right]
\qquad (i=1,2,...,M).
\label{volterra0}
\end{equation}
For large  values of  $M$, it  is reasonable  --as a  phenomenological
approach-- to choose the parameters $r_i$ and $\kappa_{ij}$ at  random
from   given   probability   distributions.   Whithin   this  type  of
representation,   the   dynamics   of   coevolving   species   can  be
characterized by statistical properties over different realizations of
parameter sets.

There are two biological systems that can potentially involve a  large
number of  coevolving populations.   The first  one is  an  ecological
system in which each population corresponds to a different  biological
species, as usually interpreted  in the theory of  population dynamics
\cite{Mikh,log}.   The  other  situation  is  a  system  in which each
population  represents  a  genotype  accessible  to  a  given  species
\cite{Schuster}.  In this situation, the number of populations can  be
sensibly larger than in the case of interacting species.  Although  in
both  cases  coevolution   is  presumably  well   described  by   Eqs.
(\ref{volterra0}), the probability distributions to be assigned to the
random  parameters  $\kappa_{ij}$,  which  represent  the  interaction
between populations,  are not  necessarily similar.   In fact,  in  an
ecological system of  several coevolving species,  mutual interactions
can  be  of  different  types  (competition,  symbiosis,  parasitism).
Within a given species, instead,  it is expected that the  interaction
is mainly competitive, as in logistic models \cite{log}.

It is well known that, in a system where many individuals compete  for
a resource, the dynamics leads to  the extinction of some of them  and
to the survival of others. This is indeed a basic fact of evolution in
the  Darwinian  sense.   Though  the  generalized Lotka-Volterra model
(\ref{volterra0}) has been  explored in detail  \cite{rusos,maylibro},
it  seems  that  a  full  characterization  --either  deterministic or
statistical--  of  the  conditions  under  which  a population becomes
extinguished  or  survives  in  the  competition  process has not been
achieved. In this paper, we  aim at analyzing this particular  problem
from a statistical viewpoint.

We consider a  large number of  coevolving species or  genotypes, each
of  them  consisting  of  a  population  of identical individuals with
density $n_i(t)$.  These populations are supposed to evolve  according
to  the  Lotka-Volterra  model  (\ref{volterra0}),  subject  to purely
competitive interactions,  i.e. with  $\kappa_{ij}\ge 0$  for any pair
$i,j$.  Since  we aim at  analyzing the statistical  properties of the
dynamics, these  coefficients will  be drawn  at random  from a  given
distribution and will remain quenched from the initial time.

For simplicity, we take $r_i =1$ $\forall$ $i$ \cite{may},  indicating
that  in  the   absence  of  competition  the  dynamics  of  all   the
populations are identical.   We are thus implicitly  identifying these
populations with the genotypes  accessible to a given  species. Within
this condition --that it is not essential to our interest and could in
fact be easily relaxed-- Eqs.(\ref{volterra0}) reduce to
\begin{equation}
\dot{n}_i(t)=n_i(t)\left[ 1-\sum_{j=1}^M\kappa _{ij}n_j(t)\right]
\qquad (i=1,2,...,M).  \label{volterra}
\end{equation}
All the coefficients $\kappa_{ij}$ will  be chosen at random from  the
same distribution $p(\kappa)$, such that $p(\kappa)=0$ for $\kappa<0$.

In the next  section we outline  the behavior of  the dynamical system
(\ref{volterra}) in phase space,  showing that its evolution  proceeds
along  a  series  of  ``pseudo-extinctions'',  in  which  some  of the
densities $n_i(t)$ can attain very low levels during long periods but,
eventually, they recover significative  values. A threshold for  these
pseudo-extinctions --that  become consequently  true extinctions--  is
suggested by the biological context of the problem. This threshold  is
introduced in our numerical study of Eqs.  (\ref{volterra}) in Section
3,  where  we  focus  the  attention  on the statistics of extinct and
surviving genotypes and try  to characterize their long-time  behavior
in terms  of their  inital conditions.   Our results  are discussed in
Section 4.

\section{Phase space structure}

The  evolution  of  the  dynamical  system  (\ref{volterra})  can   be
described  in   terms  of   a  semi-quantitative   analysis  of    the
corresponding phase-space topology, which  is determined by the  fixed
points of (\ref{volterra}) and the associated invariant manifolds.

The equations for the fix-point coordinates $n_i^*$ read
\begin{equation} \label{fix}
n_i^*\left( 1-\sum_j \kappa_{ij} n_j^*\right)=0 \ \ \ \ \ \ \
(i=1,2,...,M)
\label{fixed}
\end{equation}
and  have,  generically,  $2^M$  solutions.  In fact, each solution to
these equations can  be characterized by  the number $M'$  of non-zero
coordinates  ($M'=0,...,M$);  let  us  call  such  a  solution an {\it
$M'$-equilibrium}.   For   a   given   choice   of   the  coefficients
$\kappa_{ij}$ the number of different $M'$-equilibria is $C(M,M')=  M!
/M'!  (M-M')!$.  Therefore  --disregarding  pathological  choices   of
$\kappa_{ij}$--  the  total  number  of  fixed  points  is  $\sum_{M'}
C(M,M')=2^M$.

Since $n_i^*$ stands  for a density,  meaningful equilibria among  the
$2^M$  fixed  points  are  those  with  non-negative  coordinates.  In
Appendix  A  it  is  proven   that,  for  random  $\kappa_{ij}$,   the
probability that all the  non-zero coordinates of an  $M'$-equilibrium
($M'\neq 0$) are positive is
\begin{equation} \label{pdos}
P(M')=2^{1-M'} .
\end{equation}
We stress  that it  is essential  to this  result that $\kappa_{ij}>0$
$\forall$ $i,j$, i.e. that the system is purely competitive. For large
$M$, the  number of  equilibrium points  with non-negative coordinates
will therefore be approximately given by
\begin{equation}
\sum_{M'\neq 0}2^{1-M'}C(M,M')\approx 2\left(\frac 32\right) ^M.
\end{equation}
It  is  interesting  to  note  that,  if   $\kappa_{ij}>\kappa_{\min}$
$\forall$ $i,j$, all the non-negative equilibria will be confined to a
certain  volume  $V$  in  phase  space,  since  $\sum_i n_i^*<1/\kappa
_{\min}$.  This   volume  shrinks   rapidly  for   growing  $M$,    as
$V=\kappa_{\min}^{-M}/M!$,  and  the  density  of  equilibrium  points
--most of which are situated on the surface of $V$, where some of  the
coordinates vanish-- grows correspondingly.

The   stability   properties   of   the   fixed   points   of   system
(\ref{volterra})  can  be  fully  analyzed  in some very special cases
only.  For  instance,  as  could  be  expected  for this logistic-like
dynamical system, the $0$-equilibrium ($n_i^*=0$ $\forall$ $i$) can be
proven  to  be  always  unstable.  Moreover,  for  a  random choice of
positive  $\kappa_{ij}$,  $1$-equilibria  are  stable with probability
equal  to  $M^{-1}$.  Finally,  the  $M$-equilibrium  ($n_i^*  \neq 0$
$\forall$ $i$) is stable if $\kappa_{ij}$ is a symmetric matrix.

Though we cannot give a detailed characterization of the stability  of
all of the $M'$-equilibria, it can be argued that, for a random system
and for large $M'$ and $M$, the eigenvalues of the linearized  problem
should  follow  a  semicircular  distribution  \cite{semic}. A typical
equilibrium point is thus  linearly unstable and it  has approximately
the same number of positive and negative eigenvalues. Correspondingly,
the  number  of  unstable  and  stable  invariant  manifolds  for each
equilibrium is more or less the same. Since the mathematical structure
of system (\ref{volterra}) prevents both the divergence of orbits  and
changes of sign in the densities $n_i(t)$, the invariant manifolds  of
positive equilibria  are necessarily  bounded and  mutually connected,
defining homoclinic and heteroclinic orbits. Most of these orbits  lie
on the surface  of the volume  that contains the  positive equilibria,
where some of the densities are exactly equal to zero.

In summary, the portion of  the $M$-dimensional phase space of  system
(\ref{volterra}) meaningful to our problem is populated by a large set
of fixed points --of the order  of $(3/2)^M$ in number-- most of  them
having positive and negative  eigenvalues, i.e. being unstable.   They
are confined to a volume of order $1/M!$ and, typically, are found  on
the surface of such volume. These equilibria are highly interconnected
through invariant manifolds which lie also on that surface and connect
stable and unstable eigenvectors. The number of those manifolds should
be of order $M(3/2)^M$.

With these elements in hand, the evolution along a typical phase-space
trajectory of the dynamical system (\ref{volterra}) can be outlined as
follows.   From a  generic initial  condition, the  orbit should  soon
approach one of  the stable manifolds,  and the system  will be driven
towards the corresponding equilibrium.  It will spend some time in the
vicinity of this  equilibrium, but if  this fixed point  is not stable
(i.e, if it  has at least  one unstable manifold,  which --as we  have
argued--  is  the  typical  case)  the  orbit  will finally leave that
neighborhood, just to be drawn along one of the unstable manifolds  of
this  first  equilibrium  point  towards  another equilibrium, that is
expected to have in turn some stable and some unstable manifolds.  The
whole process will repeat itself  and the system will wander  in phase
space,  typically  visiting  the  neighborhoods  of  a large number of
unstable fixed points, until it eventually finds a stable equilibrium.
This  is  reminiscent  of  the  complex  behavior of Boolean evolution
models  on  random  landscapes  \cite{K},  which  --in  contrast  with
Lotka-Volterra models-- are however  discrete (in space and  time) and
stochastic.

We stress  that, in  wandering from  one equilibrium  to another,  the
orbit is expected to approach  more and more the successive  invariant
manifolds  that  drive  the  dynamics  of  the system \cite{MP}.  This
implies, in particular, that the  system will spend longer and  longer
periods  in  the  immediate  vicinity  of  those  equilibria.   Since,
typically,   the   equilibria   have   some   null   coordinates,  the
corresponding densities will approach a vanishing state but ---as  the
system  escapes  from  each  unstable  equilibrium  point---  they can
eventually recover appreciable values.   As the evolution proceeds,  a
given density can  therefore practically vanish  during a rather  long
time, but  can then  increase and  become again  significative in  the
whole dynamics. We shall come back to these pseudo-extinctions in  the
next sections, to  discuss their relevance  in the numerical  study of
the system and its biological interpretation.

Finally, it  is worthwhile  to remark  that the  existence of a stable
equilibrium  point,  able  to  definitively  attract  an  orbit, is in
principle  not  guaranteed.  Moreover,  even  if one or several stable
points do exist,  it is not  insured that their  bassins of attraction
cover the  whole space  of initial  conditions. The  system could thus
perform  a  chaotic  orbit  or   become  trapped  in  a  limit   cycle
\cite{MP,JCP}.

\section{Numerical analysis}

We  have  performed  an  extensive  numerical  investigation of system
(\ref{volterra}).   Each   realization  consists   of  the   numerical
integration of the equations, after a random choice of the interaction
matrix and the initial conditions. In all the results presented  here,
the interaction coefficients  $\kappa_{ij}$ have been  randomly chosen
from a uniform distribution  in the interval $[\kappa_0-\Delta  \kappa
,\kappa _0+\Delta \kappa ] $, with $\Delta \kappa\leq\kappa_0$, but we
have tested that other probability distributions --always defined  for
$\kappa>0$--  produce  essentially  the  same  results. Similarly, the
initial densities  have been  uniformly distributed  at random  in the
interval $[0,n_{\max}]$.   A proper  rescaling of  densities and  time
makes it possible  to fix, without  generality loss, $\kappa_0=1$  and
$n_{\max}=1$. The  only parameter  to vary  in these  distributions is
therefore $\Delta \kappa$. In the following we describe the  dynamical
behavior of (\ref{volterra}) as drawn from our numerical calculations.

\subsection{Pseudo-extinctions and density threshold}

In Figure 1, we show the  evolution of several typical densities in  a
system of $M  = 20$ genotypes,  for $\Delta\kappa=0.5$. Note  that, to
ease  the  appreciation  of  certain  details,  both  the time and the
density axis are logarithmic. In  the inset the same curves  are shown
in a semilogarithmic plot, with logarithmic time axis.  The phenomenon
of pseudo-extintions is  clearly seen in  some of the  curves. We have
checked  that,  in  some  realizations,  one  or  more  densities  can
temporarily attain values  as small as  $n \sim 10^{-16}$,  and then
grow to levels of the order of their initial values.  The verification
that  pseudo-extinctions  do  occur  --as  predicted,  in the previous
section, from our analysis of  the phase space structure-- points  the
attention to a further ingredient,  which is not present in  the model
as described by (\ref{volterra}), but has to be necessarily taken into
account in a  system where the  variables are, actually,  discrete. In
fact, the population  density of a  species or a  genotype confined to
certain spatial domain $\Omega$ of volume $V_\Omega$ cannot be smaller
than $V_\Omega^{-1}$, unless it vanishes. In a description in terms of
densities,  it  is  therefore  necessary  to fix a threshold \cite{T},
below which the  only value accessible  to the density  is effectively
zero.   Besides  this  biological  argument  for introducing a density
threshold,  we  must  stress  that  in our numerical calculations this
element is also necessary to avoid spurious effects of finite computer
precision on the results.  A threshold $n_0$ has been thus  introduced
as an additional  parameter in the  numerical calculations, in  such a
way  that  if  a  density  attains  a  value  lower  than  $n_0$ it is
automatically set to zero.

From the analytical viewpoint, it can be argued that the  introduction
of   a   threshold   changes   the   stability   of   almost  all  the
$M'$-equilibria. Roughly speaking, whereas without threshold an  orbit
could  approach  an  equilibrium  following  a stable manifold just to
leave it along an unstable one, with threshold the system can  instead
be ``captured''  by the  equilibrium point  if the  orbit crosses  the
threshold.  What  was  an  unstable  equilibrium becomes, in effective
terms, a stable one.

In order  to illustrate  the different  behavior of  systems with  and
without density threshold, we have chosen to analyze the evolution  of
the total density
\begin{equation}
N(t)=\sum_{i=1}^Mn_i(t),
\end{equation}
as a  global characterization  of the  phase space  dynamics. Figure 2
displays the evolution  of $N(t)$ for  two systems of  100 species. In
both cases the interaction is defined by $\Delta\kappa=1$. One of  the
curves corresponds to  the system without  threshold (besides  that
imposed by  the smallest  representable number  in the  computer). The
other  one  corresponds  to  the  same  system --with the same initial
conditions and interaction matrix--  with a threshold $n_0  =10^{-6}$.
Note that the time scale is again logarithmic. It can be seen that, in
this  realization,  the  orbit  of   the  first  system  follows   the
qualitative behavior described in the previous section. It passes near
some equilibria --where $N(t)$ remains practically constant-- spending
exponentially larger and  larger times in  their neighborhood. When  a
threshold is present, the system is always attracted to one of the new
``stable'' equilibria.  In the case of Fig. 2, the system is  captured
at  $t\approx  100$  by  an  equilibrium  point  that,  although being
unstable in the first case, acts now as a stable fixed point for  this
orbit.

\subsection{Statistics of survivals}

According  to  our  simulations,  the  main  feature  in the long-time
dynamics of system (\ref{volterra})  --with or without threshold--  is
that it evolves to a situation  in which most of the densities  vanish
--at finite times if $n_0\neq  0$ or asymptotically if $n_0=0$.   This
behavior can be  identified with the  extinction of the  corresponding
genotypes.   In any  case, for  a large  system, a  variable number of
surviving populations is found at long  times. In Fig. 3, we show  the
distribution  of  the  number  of  surviving populations for different
values of $\Delta k$. Each  curve was constructed from the  results of
2000  realizations  in  a  system  with $M=100$ and $n_0=10^{-6}$. The
final time in each realization  was chosen so that a  stationary state
had  been  reached,  which  was  checked  to  be  a  solution  to Eqs.
(\ref{fix}).

The distribution of the number of surviving populations is generally a
bell-shaped curve, its width and maximum depending on the  probability
distribution $p(\kappa)$. It  can be seen  in Fig. 3  that for $\Delta
\kappa$ small  enough the  curve is  relatively broad,  and that wider
interactions reduce the overall stability of the system, leading to  a
shift of the curve towards a situation where less species survive. The
correlation between the maximum  of the distribution of  survivors and
the width of $p(\kappa  )$ is shown in  the inset, in a  log-log plot.
Observe that for the smallest value, $\Delta\kappa=0.02$, the  maximum
of the distribution coincides with the total number of species in  the
system.

It is worthwhile to note that the probability of having a non-negative
$M'$-equilibrium, $P(M')=2^{1-M'}C(M,M')$ (cf.  Section 2), is  also a
bell-shaped curve as a function  of $M'$. This fact should  however be
considered only as indicative of the profile of the curves in Fig.  3.
In  fact,  the  probability  that  an initial condition approaches any
$M'$-equilibrium depends not only on their number for a given $M'$ but
also  on  the  size  of  their  bassins of attraction, about which our
phase-space analysis provides no information.

\subsection{Characterization of survivors}

A very  natural question  arises now,  yet not  easy to  answer: which
populations survive? Are they characterized by some particular initial
condition, by some peculiarity  in the interaction with  the remaining
populations? In  the study  of a  real ecological  system, it would be
desirable to give an answer to these questions in terms of  quantities
accessible  from  observations.   It  is  thus reasonable to consider,
since they are probably the easiest to measure, the initial  densities
and their  initial time  derivatives. Besides  being accessible, these
quantities characterize the  initial interactive scenario:   according
to Eqs.   (\ref{volterra}), the density  measures the effects  of each
population on itself, while its first time derivative accounts for the
influence of the remaining species.

We have found that an  answer based on deterministic arguments  cannot
be given to such questions. According to the statistics collected from
the  simulations,  we  conclude  that  only  a weak correlation exists
between survival and the initial conditions.  This correlation can  be
evidenced  by  calculating  the  distribution  of final densities as a
function of the initial one, providing thus a probabilistic answer  to
those  questions.   In  Figs.  4  and  5  we  show  (full  lines)  the
distribution of survivors as functions of initial values, for  general
asymmetric systems ($\kappa_{ij}  \neq \kappa_{ji}$).   Fig.  4  shows
that the probability of survival is almost uniform in the whole  range
of initial densities,  with a slighty  higher probability of  survival
for the largest ones.  In Fig.  5 an associated distribution is shown:
the number of survivors as a function of the initial derivative of the
density. Here, an enhancement of the probability of surviving is  seen
around a  relatively large  (and negative)  value of  the initial time
derivative.

We have also found that  this correlation between the final  state and
the initial condition is stronger in symmetric systems ($\kappa_{ij} =
\kappa_{ji}$).   In Figs.  4 and  5, the  same functions  are shown as
dashed  lines  for  symmetric  systems.   In  this  case,  there is an
enhancement  in  the  probability  of  survival for those species that
start with a higher initial population.  Although considering a purely
symmetric interaction matrix is  irrelevant from the biological  point
of  view,  these  results  suggest  that  the  statistical correlation
between initial and final states --and, in particular, between initial
conditions and survival  probability-- can depend  in a rather  strong
manner on additional constraints in the interaction matrix.

\section{Conclusions}

We have analyzed a dynamical system  that represents the  evolution of
many species  coupled by  Lotka-Volterra interactions.  The study  has
been  restricted   to  systems   where  the   interaction  is   purely
competitive.  This  dynamical  system  describes,  in  principle,  two
different biological  systems. The  first is  an idealized  ecological
system of interacting species.  To represent more realistic ecological
systems,   the   connectivity   of   the   model   should   be   built
correspondingly,  typically,   with  several   levels  of   preys  and
predators.

On the other hand, the model can also describe the system of genotypes
present in, or accessible to, a single species or population. Within a
single species, the number of  competing genotypes can be much  larger
than  the  number  of  competing  species  in  an ecological niche. Of
course, not  all of  them strive  and result  finally expressed in the
living population, and this is precisely the problem we have addressed
in this work.

We have  found that  the evolution  of the  system follows complicated
orbits  in  phase  space.  These  orbits  drive  the  system  from the
neighborhood of one of the  many equilibria to another, regardless  of
their stability.  Systems with a finite population threshold, that may
represent  more  accurately  real  biological systems, eventually fall
into  a  stable  equilibrium  situation.  As  time elapses, a variable
number of populations become extinct through the interaction with  the
others.  In general, more  than one species survive, in  contrast with
the ``principle of competitive  exclusion'' \cite{log} (that is  known
to be of limited validity).  The number of surviving ones --those that
finally  reach  equilibrium--  is   characterized  by  a   bell-shaped
distribution, whose width  and maximum depend  on the distribution  of
interactions.

Besides  competition,  a  population  of  genotypes is also subject to
changes that arise from random mutations and recombination during  the
reproduction of the organisms \cite{Eigen}. The description of such  a
system would require a modification of model, whose behavior cannot be
predicted {\it a priori}. Mutations  can be easily taken into  account
by allowing a new  interaction, namely random transitions  between the
genotypes \cite{feistel}. The analysis  of this system is  the subject
of work in progress.

\section*{Acknowledgements}

The  authors  are  grateful  to  D.E.  Strier and to S.C. Manrubia for
enlightening  discussions.  D.H.Z.  thanks  the Alexander von Humboldt
Stiftung  for  his  fellowship  and  the  Fritz  Haber  Institut   for
hospitality  during  his  stay  in  Berlin.  Financial  support   from
Fundaci\'on Antorchas, Argentina, is also acknowledged.

\appendix

\section{}

\noindent

{\bf  Theorem.}  Let  $\{  \kappa_{ij}  \}$  be  an $M\times M$ random
matrix,  whose  coefficients  are  drawn  from  the  same  probability
distribution $p(\kappa)$, with $p(\kappa)=0$ for $\kappa \le 0$. Then,
the probability that the solution to the set of linear equations
\begin{equation}    \label{A.1}
\sum_{j=1}^M \kappa_{ij}n_j =1, \ \ \ \ \ \ (i=1,2,...,M),
\end{equation}
has positive components, $n_i>0$ $\forall$ $i=1,2,...,M$, is
$P=2^{1-M}$.

\noindent
{\it  Proof.}  For  $M=1$,  $n_1  =  1/\kappa_{11}$,  which  is always
positive ($P=1$). For $M=2$, the  system can be explicity solved  and,
in particular, we get
\begin{equation}    \label{A.2}
{n_1\over n_2}= {\kappa_{22}-\kappa_{12}\over
\kappa_{11}-\kappa_{21}}.
\end{equation}
The  symmetry  of  this  expression  with  respect to the coefficients
$\kappa_{ij}$  makes  clear  that  $n_1/n_2>0$ with probability $1/2$,
irrespectively of the form of $p(\kappa)$. Now --since $\kappa_{ij}>0$
$\forall$ $i,j$-- if  $n_1$ and $n_2$  have the same  sign and satisfy
Eqs.  (\ref{A.1})  with  $M=2$,  they  must  be  positive.  Therefore,
$P=1/2$.

For $M>2$, we take any pair   of equations from (\ref{A.1}) --say  the
$k$-th and the $l$-th-- and rewrite them as
\begin{equation}    \label{A.3}
\begin{array}{rl}
\kappa_{kk}n_k +\kappa_{kl}n_l &= 1-\sum_{j\neq k,l} \kappa_{kj} n_j,
\\
\kappa_{lk}n_k+\kappa_{ll} n_l &= 1-\sum_{j\neq k,l} \kappa_{lj} n_j.
\end{array}
\end{equation}
This can also be put in the form
\begin{equation}    \label{A.4}
\begin{array}{rl}
\kappa_{kk}'n_k +\kappa_{kl}'n_l &= 1, \\
\kappa_{lk}'n_k+\kappa_{ll}' n_l &= 1,
\end{array}
\end{equation}
with $\kappa_{kk}'=\kappa_{kk}/(1-\sum_{j\neq k,l} \kappa_{kj}  n_j)$,
and  analogous  expressions  for  $\kappa_{kl}'$,  $\kappa_{lk}'$  and
$\kappa_{ll}'$.

Note that  for fixed,  arbitrary values  of $n_j$  ($j\neq k,l$),  the
functional form  of the  primed coefficents  in terms  of the original
ones is  the same.  The probability  distributions for $\kappa_{kk}'$,
$\kappa_{kl}'$,  $\kappa_{lk}'$  and   $\kappa_{ll}'$  are   therefore
identical. Hence, as a consequence  of the previous result for  $M=2$,
the probability that  $n_k/n_l$ is positive  equals $1/2$. This  holds
irrespectively of the distribution  for the primed coefficients,  i.e.
irrespectively of the values of $n_j$ ($j\neq k,l$). The relative sign
of  any  two  components,  $n_k$  and  $n_l$,  of the solution to Eqs.
(\ref{A.1}) is  then statistically  independent of  the values  of the
other components.

To insure  the positivity  of all  the components  it is sufficient to
consider  $M-1$  ratios  $n_k/n_l$,   for  instance,  $n_1/n_l$   with
$l=2,3,...,M$. According  to the  above results,  the probability that
all  these  ratios   are  positive  is   $(1/2)^{M-1}$.  Now   --since
$\kappa_{ij}>0$ $\forall$ $i,j$-- if all the $n_i$ have the same  sign
and satisfy Eqs.  (\ref{A.1}),  they must be positive. Therefore,  $P=
(1/2)^{M-1}=2^{1-M}$.

\newpage

\begin{figure}
\scalebox{.5}{\rotatebox{270}{\includegraphics{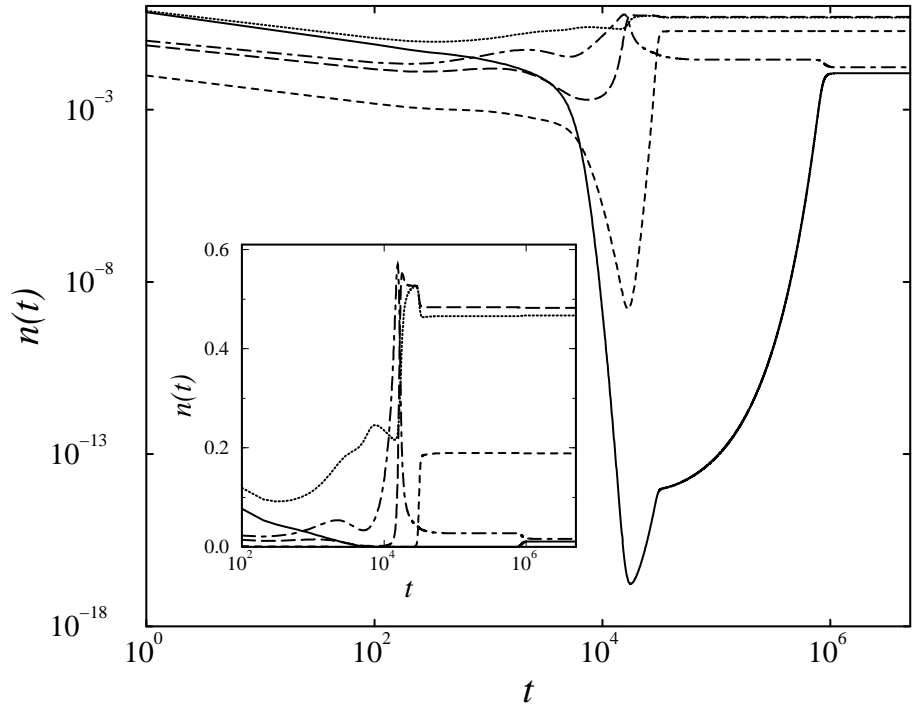}}}
\caption{
Time evolution of   the density of some selected  populations,
from  a  system   consisting   of  20   populations.  Both  axis   are
logarithmic, to emphasize how some of the populations, after  becoming
almost extinct, grow again to  significative values.  Inset: the  same
curves, in a log-linear plot.
}
\end{figure}

\newpage
\begin{figure}
\scalebox{.5}{\rotatebox{270}{\includegraphics{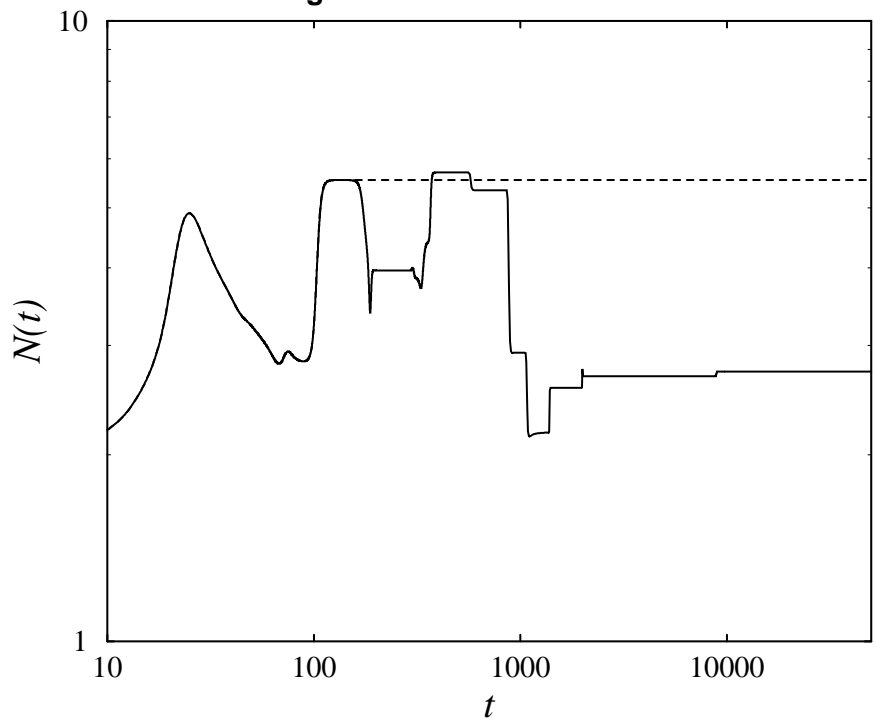}}}
\caption{
Time  evolution  of  the  total  density.  Full line: system
without density threshold. Dashed  line: the same system  with density
threshold. Note the logarithmic scale in the time axis.
}
\end{figure}

\newpage
\begin{figure}
\scalebox{.5}{\rotatebox{270}{\includegraphics{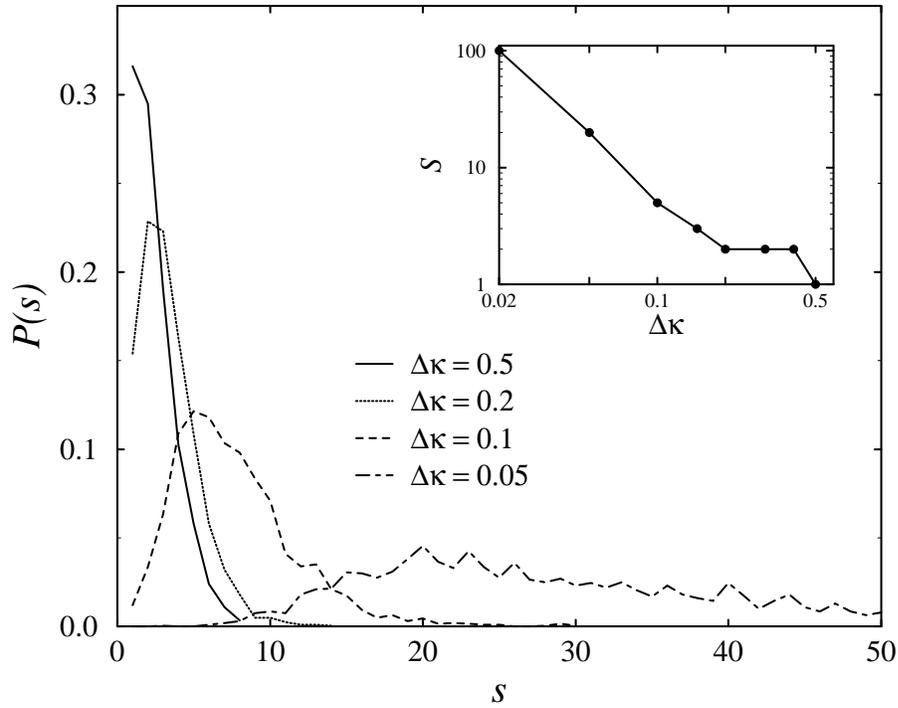}}}
\caption{
Distribution  of the number  of survivors, $P(s)$.  Each curve
corresponds to  2000 realizations  of systems  with $\Delta\kappa$  as
shown  in  the  legend.  Inset:  The  position  of  the maximum of the
distribution, $S$, as  a function of  the width $\Delta\kappa$  of the
distribution $p(\kappa )$.
}
\end{figure}

\newpage
\begin{figure}
\scalebox{.5}{\rotatebox{270}{\includegraphics{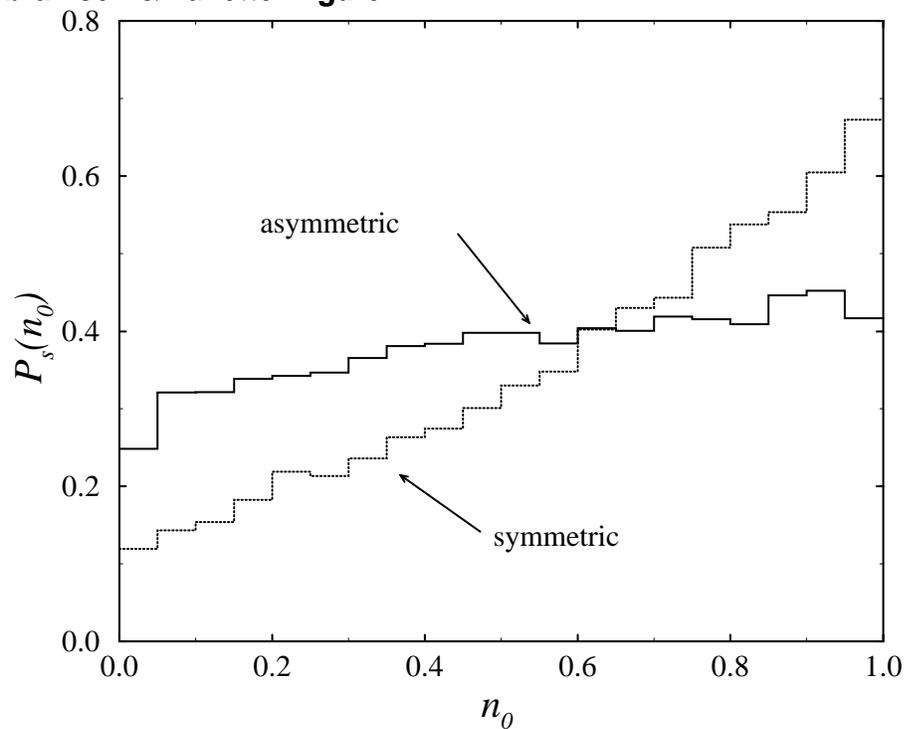}}}
\caption{
The probability of survival  of a single species, $P_s$, as  a
function of its initial density,  $n_0$. The two curves correspond  to
2000 realizations of a symmetric and an asymmetric system.}
\end{figure}

\newpage
\begin{figure}
\scalebox{.5}{\rotatebox{270}{\includegraphics{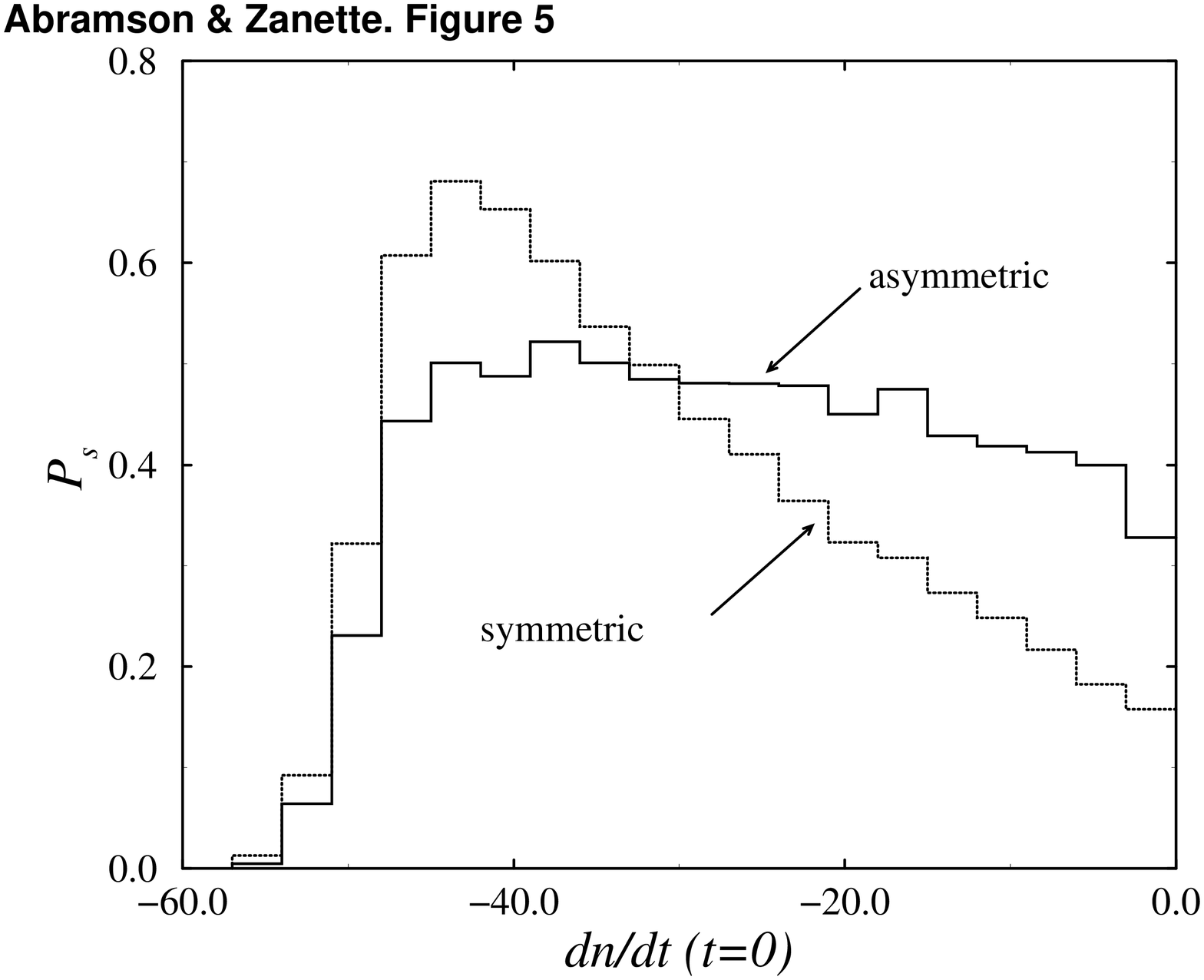}}}
\caption{
The probability  of survival,  $P_s$   as a  function of  the
initial  derivative  of  the  density,  $dn/dt|_{t=0}$. The two curves
correspond  to  2000  realizations  of  a  symmetric and an asymmetric
system.}
\end{figure}


\begin{thebibliography}{9}

\bibitem{evol} P. Skelton, {\it Evolution: A Biological and
Paleontological Approach} (Addison-Wesley, Wokingham, 1994).

\bibitem{may}  R.M. May, Nature {\bf 238}, 413 (1972).

\bibitem{rusos}  Yu.M. Svirezhev and D.O. Logofet, {\it Stability of
Biological Communities} (Mir, Moscow, 1983).

\bibitem{Mikh} A.S. Mikhailov, {\it Foundations of Synergetics I}
(Springer, Berlin, 1990) Chap. 7.

\bibitem{log} J.D. Murray, {\it Mathematical Biology}
(Springer-Verlag, Berlin, 1993).

\bibitem{Schuster} P. Schuster, Physica {\bf 22D}, 100 (1986).

\bibitem{maylibro}  R.M. May, {\it Stability and Complexity in Model
Ecosystems} (Princeton University Press, 1974).

\bibitem{semic} C.E. Porter, {\it Statistical Theories of Spectra:
Fluctuations} (Academic Press, New York, 1965).

\bibitem{K}S.A. Kauffman, Physica D {\bf 42}, 135 (1990);
J. Theor. Biol. {\bf 149}, 467 (1991).

\bibitem{MP}A.S. Mikhailov and I.Yu. Poteryaiko, Physica D {\bf 53},
13 (1991).

\bibitem{JCP} P.E. Phillipson and P. Schuster, J. Chem. Phys. {\bf
79}, 3807 (1983).

\bibitem{T} L.S. Tsimring, H. Levine and D.A. Kessler, Phys. Rev.
Lett. {\bf 76}, 4440 (1996).

\bibitem{Eigen} M. Eigen, Naturwissenschaften {\bf 58}, 465 (1971).

\bibitem{feistel}  T. Boseniuk, W. Ebeling and A. Engel, Phys. Lett.
{\bf 125 A}, 307 (1987).

\end{thebibliography}
\end{document}